\documentclass[preprint,showkeys, preprintnumbers,amssymb]{revtex4}
\usepackage{graphicx}
\usepackage{amssymb}
\usepackage{amsmath}
\usepackage{bm}
\begin{document}
\setcounter{page}{1}
 \title{An ac field probe for the magnetic ordering of magnets with random anisotropy}

\author{Ha M. Nguyen}
\thanks{Author to whom correspondence should be addressed. Electronic mail: nmha@ess.nthu.edu.tw}
\author{Pai-Yi Hsiao}
\thanks{Electronic mail: pyhsiao@ess.nthu.edu.tw}
\affiliation{Department of Engineering and System Science, National Tsing Hua University, Hsinchu 30013, Taiwan }
\date{\today}

\newpage
\begin{abstract}
A Monte Carlo simulation is carried out to investigate the magnetic
ordering in magnets with random anisotropy (RA). Our results show
peculiar similarities to recent experiments that the real part of ac
susceptibility presents two peaks for weak RA and only one for
strong RA regardless of glassy critical dynamics manifested for
them. We demonstrate that the thermodynamic nature of the
low-temperature peak is a ferromagnetic-like dynamic phase
transition to quasi-long range order (QLRO) for the former. Our
simulation, therefore, is able to be incorporated with the
experiments to help clarify the existence of the QLRO theoretically
predicted so far.
\end{abstract}

 \pacs{}
\keywords{ Dynamic transition, random magnetic anisotropy, amorphous
magnets, Monte Carlo simulation, critical slowing down, complex
susceptibility}

\maketitle

Over decades, great interest has been intensively addressed to
rare-earth magnetic glasses of random magnetic anisotropy (RMA).
However, the nature of magnetic phase transition (MPT) and magnetic
ordering in such random magnets has still been far from being
completely understood \cite{ref1}.  First, Monte Carlo (MC)
simulations were opposite to the prediction of renormalization group
theories \cite{ref2,ref3} to reveal that the second-order MPT exists
in three-dimensional weak RMA systems of XY \cite{ref4} and
Heisenberg spins \cite{ref5,ref6}. Second, Itakura and Arakawa
\cite{ref9} have demonstrated that a crucial additional vortex
energy should be included in the Imry-Ma type arguments, which have
predicted the absence of ferromagnetic long range order (LRO) in
magnets of random field (RF) \cite{ref7} and RMA \cite{ref8} for
space dimensions $d<4$, to explain the power-law correlation of
quasi-long range order (QLRO) in the Bragg glass state of impure
superconductors \cite{ref10}, and showed MC results of the power-law
scenario for the weak RF model of XY spins. Feldman \cite{ref11} has
theoretically shown that QLRO can emerge instead of LRO in $d = 4
-\epsilon$ dimensions and is common in such impure systems of
continuous non-Abelian symmetry as magnets of weak RF and RMA. In
addition, QLRO has  also been clearly evidenced in a number of MC
simulations \cite{ref5,ref9,ref12}, of which a power-law spin
correlation function has been found to indicate a ground state of
QLRO in weak RMA systems of Heisenberg spins \cite{ref5}. In spite
of these theoretical conjectures, the lack of direct experimental
and theoretical agreements in the literature leads to the questions
of (i) whether the magnetic transition and low-temperature magnetic
order in weak RMA magnets are ferromagnetic-like, and (ii) whether
the so-called RMA model \cite{ref13} can be applied to understand
such phenomena in real materials.

In this {\it Letter}, we address these questions by conducting a MC
simulation upon the RMA model \cite{ref13}. Our simulation aims to
clarify an experimental possibility that the singularity on the
temperature-dependent curves of the real part of the ac
susceptibility, $\chi^{\prime}(T,\omega)$, for a  weak RMA glass of
a-Ho$_{28}$Fe$_{72}$ amorphous film reported by  Saito {\it et al.}
\cite{ref14} manifests a second-order MPT, which is discriminated in
nature from the magnetic glassy phase transition (GPT) in strong RMA
glasses, for instance, Dy$_{40}$Al$_{24}$Co$_{20}$Y$_{11}$Zr$_{5}$
bulk glass \cite{ref15}. Differing from the MPT, the GPT is
indicated by a glassy critical slowing down law without any
singularity shown on $\chi^{\prime}(T,\omega)$ at the transition
temperature, $T_{\rm g}$ \cite{ref15}. Notice that a similar
scenario, but for a case of RF systems, has been existed in the
literature when Schremmer and Kleemann \cite{ref16} demonstrated for
an orientational glass system of K$_{1-x}$Li$_{x}$TaO$_{3}$ with $x
= 0.063$ (a doping well above the glassy and ferroelectric boundary
$x_{\rm c} \sim 0.022$) that the singularity on the real part of the
ac dielectric permittivity, $\epsilon^{\prime}(T,\omega)$, is of a
transition to the long-range ferroelectric phase. However, this is a
first-order transition.

\begin{figure}
       \begin{center}

         \resizebox{130 mm}{!}{\includegraphics{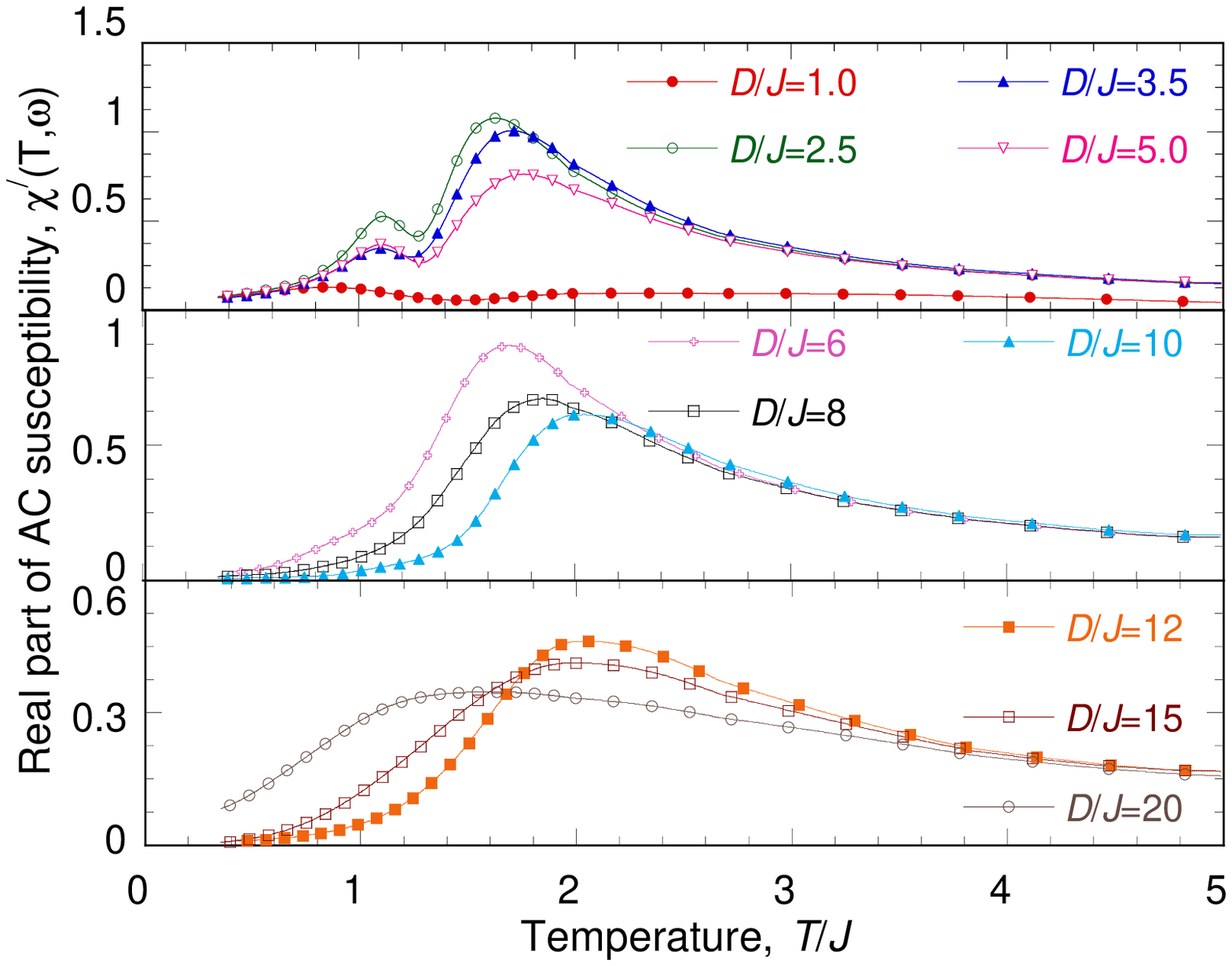}}
       \end{center}
           \caption{$\chi^{\prime}(T,\omega)$ curves at $\omega=3\times10^{-3}$
            and $H_{0}=0.05$ for various $D/J$ values. $\chi^{\prime}(T,\omega)$ exhibits  distinct characteristics of
            weak ($D/J \le 5$)
            and strong ($D/J > 5$) degrees of RMA.}
        \label{fig1}
\end{figure}

In the light of these experiments, we show in the present work that
the nature of the aforementioned singularity for the weak RMA magnet
of  a-Ho$_{28}$Fe$_{72}$ amorphous film \cite{ref14} can be
understood dynamically with the concept of {\it dynamic transition}
within the framework of the RMA model of three-dimensional
Heisenberg spins \cite {ref13}, of which the Hamiltonian can be
written as
\begin{equation}
\mathcal{H}=-J\sum_{\langle i,j\rangle}{\vec{S}_{i}\cdot
\vec{S}_{j}}-D\sum_{i}{(\hat{a}_{i}\cdot \vec{S}_{i})^{2}}
-H\sum_{i}{\vec{S}_{i}\cdot \hat{z}},
\label{eq1}
\end{equation}
where the first term is due to the exchange coupling with strength
$J>0$ between nearest-neighbor spins, the second is for on-site RMA
with strength $D>0$, and the last is the Zeeman term with the
presence of an external field of strength $H$ along $\hat{z}$ axis.
$\vec{S}_{i}$ and $\hat{a}_{i}$ are unit vectors representing the
spin (an annealed variable) and the random easy axis (a
randomly-quenched variable) at site $i$, respectively. In Eq.
(\ref{eq1}), the anisotropy to exchange ratio, $D/J$, plays the role
of the degree of RMA. One prominent effect of the degree of RMA is
clearly observed in Fig. \ref{fig1}. Here, $\chi^{\prime}(T,\omega)$
is simulated using the same MC technique as that described in our
previous papers \cite{ref17,ref18} for simple-cubic-lattice systems
of $L\times L\times L$ ($L=10$) Heisenberg spins as an external ac
field, $H=H_{0}\sin(\omega t)$, is applied, where $H_{0}/J=0.05$,
time $t$ is in MC step (MCS), and frequency $\omega$ is in
MCS$^{-1}$. Each data point is averaged over $50$ realizations of
\{$\hat{a}_{i}$\}$^{L^{3}}_{1}$.  As shown in Fig. \ref{fig1},
curves of $\chi^{\prime}(T,\omega)$ with $\omega =3\times10^{-3}$
exhibit two peaks for small values of $D/J$, i.e., weak RMA ($D/J\le
5$). The high-temperature peak is responsible for an Arrhenius-type
relaxation which is in common with that for those curves of large
values of $D/J$ of strong RMA, whereas the low-temperature peak
peculiarly characterizes another magnetic nature of magnets of weak
RMA, the position of which is almost insensitive to the change of
anisotropy strength. Notice that in our simulation, we mimic the
measurement protocol that the system is cooled in the ac field to
the lowest temperature then carrying out the calculation of ac
susceptibility and other quantities when heating the system up. The
reason for this choice is because we have seen in our simulation
that, unlike the RF system of K$_{1-x}$Li$_{x}$TaO$_{3}$ with $x =
0.063$ \cite{ref16}, cooling the systems of weak RMA in a nonzero dc
field even as small as $H_{0}$ shall unexpectedly result in the
suppression of the low-temperature peak of $\chi^{\prime}(T,\omega)$
curve and the curve looks like that of strong RMA, i.e., an one-peak
curve. Interestingly, these distinct characteristics of
$\chi^{\prime}(T,\omega)$ for weak and strong RMA systems are
consistent with results reported by Itakura \cite{ref5} that the
function $G(r)\propto r^{-\eta-1}\exp(-r/\xi)$ can be used to
describe spin correlation of the ground state for the RMA model in
Eq. (\ref{eq1}). The correlation length $\xi$ is finite for large
values of $D/J$ while it is infinite for weak RMA of $D/J \le 5$ so
that the spin correlation reduces to a frozen power law of QLRO
ground state, $G(r)\propto r^{-\eta-1}$. We remark that we shall
only focus on a weak RMA glass of $D/J=3.5$ and a strong RMA glass
of $D/J=10$ which are typical of the RMA model of Heisenberg spins
in Eq. (\ref{eq1}) to understand magnetic behaviors of weak RMA
a-Ho$_{28}$Fe$_{72}$ \cite{ref14} and strong RMA
Dy$_{40}$Al$_{24}$Co$_{20}$Y$_{11}$Zr$_{5}$ glasses \cite{ref15},
respectively.

\begin{figure}
        \begin{center}
          \resizebox{130 mm}{!}{\includegraphics{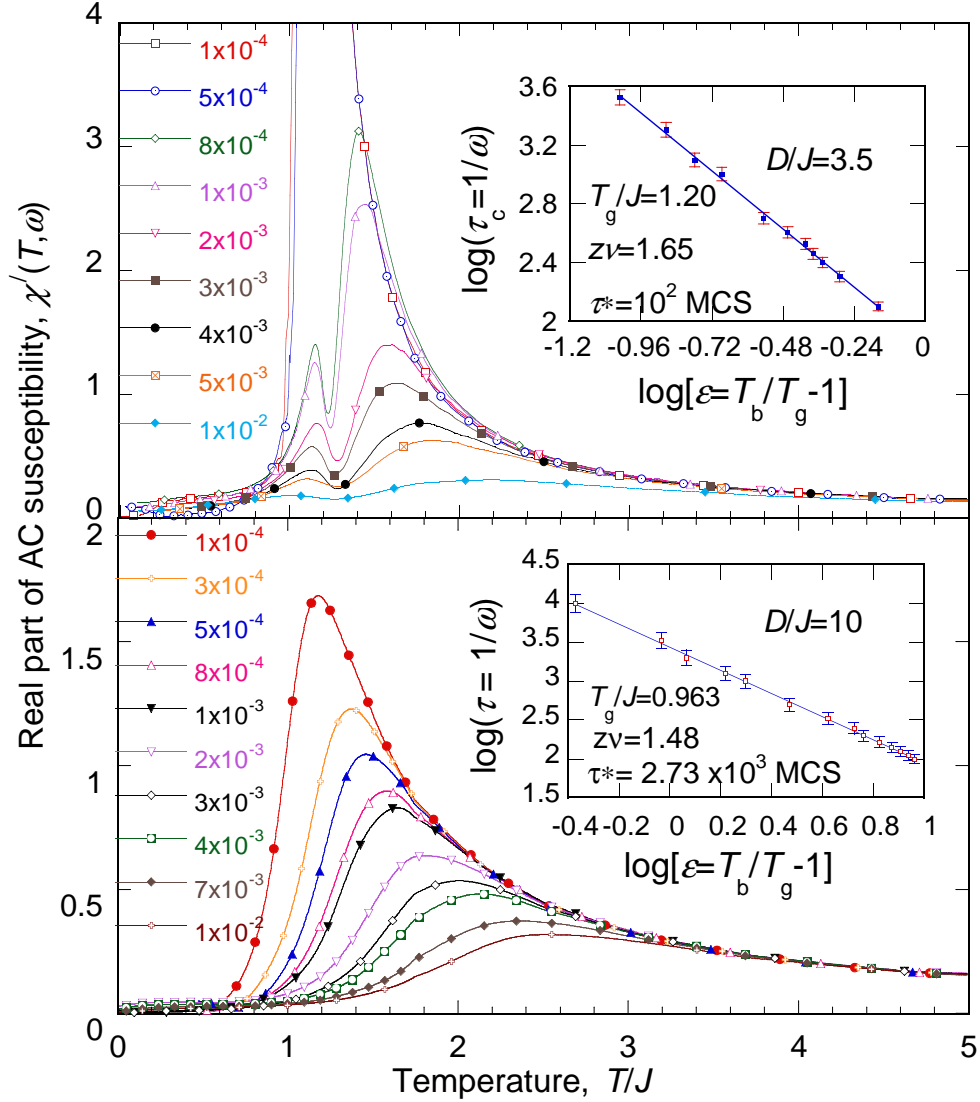}}
        \end{center}
           \caption{Temperature dependence of $\chi^{\prime}(T,\omega)$ at various
            frequencies and $H_{0}=0.05$ for $D/J=3.5$ (upper panel) and $D/J=10$ (lower panel).
            Insets show laws of critical slowing-down dynamics for the two cases, respectively.}
        \label{fig2}
\end{figure}

Figure \ref{fig2} presents the temperature dependence of
$\chi^{\prime}(T,\omega)$ at different frequencies and
$H_{0}/J=0.05$ for $D/J=3.5$ and $D/J=10$ cases. For $D/J=3.5$,
curves of $\chi^{\prime}(T,\omega)$ exhibit two peaks. The position
 of the low-temperature one, $T_{\rm{p}}$, is
insensitive to frequency and is at about $T_{\rm p}/J\approx 1.15$.
The position of the high-temperature one, $T_{\rm b}(\omega)$,
shifts toward low temperature in an Arrhenius way with decreasing
frequency in addition to increasing the heights of the two peaks. At
sufficient low frequencies, in this case $\omega \le
5\times10^{-4}$, the two peaks merge together so that
$\chi^{\prime}(T,\omega)$ rockets up then drops abruptly at about
$T_{\rm c}/J=1.0$. This fashion is what has been experimentally
shown for the a-Ho$_{28}$Fe$_{72}$ glass and the ``dip" in
$\chi^{\prime}(T,\omega)$, which occurs at the same position of the
single peak in $\chi^{\prime\prime}(T,\omega)$ (not shown),
indicates a signature of the MPT singularity \cite{ref14}. Besides,
the temperature dependence of the ac susceptibility obtained in our
MC simulations for the $D/J=3.5$ case also shows another feature
resembling the experiment of a-Ho$_{28}$Fe$_{72}$ glass. In contrast
to spin glasses (SGs), $\chi^{\prime\prime}(\omega) >
\chi^{\prime}(\omega)$ in the vicinity of the singularity, which,
according to Saito {\it et al.} \cite{ref14}, ``implies that the
center $\tau_{\rm c}$ of distribution of relaxation time $g(\ln
\tau)$ is much longer than the measuring time constant
$t=1/\omega$.'' Focussing on the dynamic behavior at the transition
region, the authors applied a phenomenological Cole-Cole model of
polydispersive relaxation which yields the ac susceptibility as
$\chi(\omega)=\chi_{\rm{a}}+(\chi_{0}-\chi_{\rm{a}})/\{t
1+(i\omega\tau_{\rm c})^{\beta}\}$ and
$g(\ln\tau)=\sin(\beta\pi)/2\pi\{\cosh[\beta\ln(\tau/\tau_{\rm{c}})]+\cos(\beta\pi)\}$,
where $\chi_{0}$ and $\chi_{\rm{a}}$ are static and high frequency
limit susceptibilities, and $0 < \beta <1$. They found that $g(\ln
\tau)$ is almost Gaussian in $\ln \tau$ and symmetric about $\ln
\tau_{\rm c}$, $\beta$ reduces from $1$ to $0.4$ and $\tau_{\rm{c}}$
becomes longer and longer with decreasing temperature toward
$T_{\rm{c}}$ in company with broadening of $g(\ln \tau)$. All of
these features are similar to SGs, however, $\tau_{\rm{c}}$ for the
a-Ho$_{28}$Fe$_{72}$ glass is several orders of magnitude longer
than those of SGs. On the other hand, the low-temperature peak is
suppressed for all frequencies in the $D/J=10$ case like that of
strong RMA of Dy$_{40}$Al$_{24}$Co$_{20}$Y$_{11}$Zr$_{5}$ glasses
\cite{ref14,ref15}. Nonetheless, we did find that there is a
well-determined transition temperature, $T_{\rm g}$, of the GPT for
both $D/J=3.5$ and $D/J=10$ cases by means of the scaling law of
critical slowing-down dynamics, $\tau_{\rm c}=\tau^{\ast}[T_{\rm
b}(\omega))/T_{\rm g}-1]^{-z\nu}$, shown in the insets of Fig.
\ref{fig2}. In terms of this scaling law, the magnetic glassy
behaviors for systems of weak and strong RMA are expected to be the
same and like those of SGs \cite{ref11}. For instance, if the
critical exponent $\nu$ of the correlation length roughly takes
values in the range of $0.7\sim0.8$ \cite{ref4,ref6} then the
dynamical exponent $z$ may be $1.85\sim2.35$, i.e., consistent with
the magnitude of those for SGs \cite{ref19}. Another example is that
Billoni {\it et al.} \cite{ref20} have reported aging phenomena for
the $D/J=3.5$ case similar to those of Heisenberg SGs at low
temperatures. To this end, a question remaining unsolved is what is
the nature of the low-temperature peak in $\chi^{\prime}(T,\omega)$
for the $D/J=3.5$ case, which will be cleared up as below.

\begin{figure}
        \begin{center}
         \resizebox{130 mm}{!}{\includegraphics{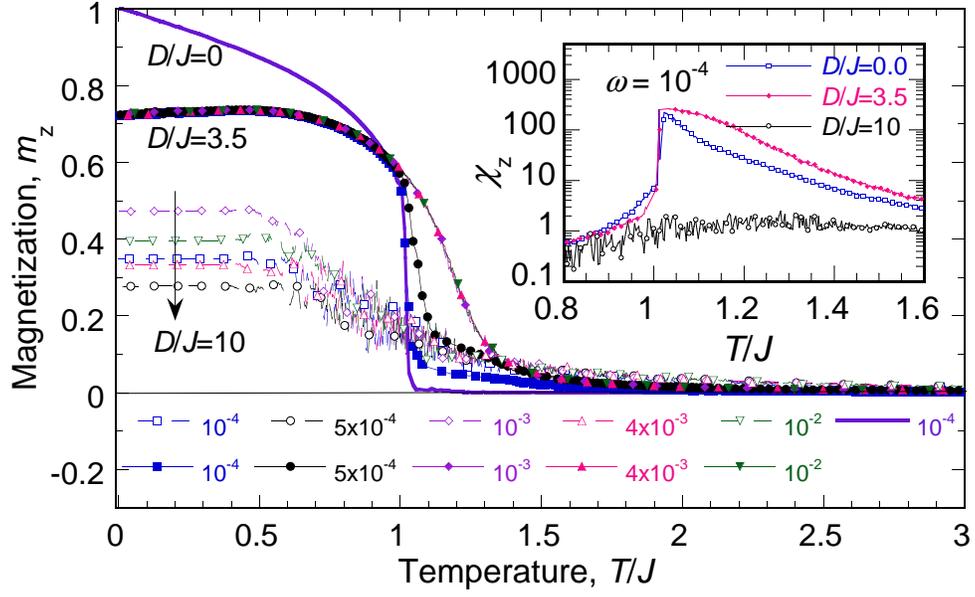}}
        \end{center}
           \caption{Temperature dependence of $m_{z}(T,\omega)$ at various
            frequencies and $H_{0}=0.05$
            for $D/J=0$ (solid line), $D/J=3.5$ (solid symbols), and $D/J=10$ (open symbols). The inset shows $\chi_{z}(T,
            \omega)$ for these cases at $\omega=1\times10^{-4}$.}
        \label{fig3}
\end{figure}

\begin{figure}
        \begin{center}
         \resizebox{130 mm}{!}{\includegraphics{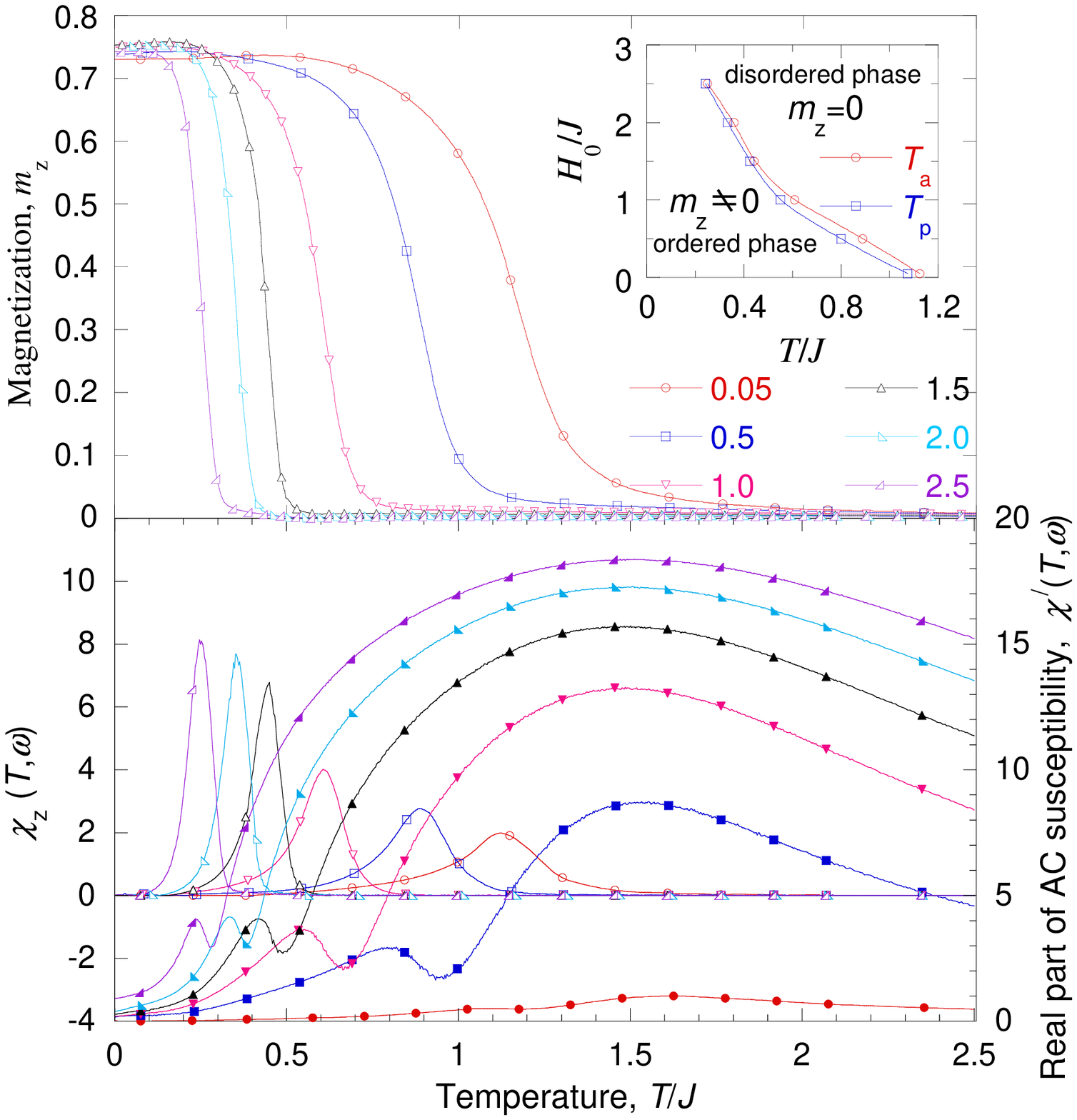}}
         \end{center}
           \caption{Temperature dependence of $m_{z}(T,\omega)$, $\chi_{z}(T,\omega)$, and
$\chi^{\prime}(T,\omega)$ for $D/J=3.5$  at $\omega=3\times10^{-3}$
and different values of $H_{0}$ (different colored curves). The
one-peak and two-peak curves are $\chi_{z}(T,\omega)$ and
$\chi^{\prime}(T,\omega)$, respectively. The inset presents the
dynamic phase diagram in $H_{0}-T$ plane, where the $T_{\rm a}$ line
(or the $T_{\rm p}$ line) shows the dynamic phase boundary between
the dynamically-ordered phase ($m_{z}\neq 0$) and
dynamically-disordered phase ($m_{z}= 0$). }
        \label{fig4}
\end{figure}

Figs. \ref{fig3} and \ref{fig4} present the results of the
temperature dependence of $m_{z}(T,\omega)$, $\chi_{z}(T,\omega)$,
and $\chi^{\prime}(T,\omega)$ from which one can see clearly
evidences of MPT for the $D/J=3.5$ case but not for the $D/J=10$
case. $m_{z}(T,\omega)$ is the averaged magnetization per spin
projected along $z$ direction, and $\chi_{z}(T,\omega)$ is the
thermodynamic fluctuation of the magnetization. In Fig. \ref{fig3},
curves of $m_{z}(T,\omega)$ are shown for these two cases with
frequencies $1\times10^{-4}\leq \omega \leq 1\times 10^{-2}$ and
$H_{0}/J=0.05$. For the sake of reference, one curve of
$m_{z}(T,\omega)$ at $\omega = 1\times10^{-4}$ (i.e., the
violet-colored solid line) is also plotted for $D/J=0$, the case of
non-anisotropic pure Heisenberg model possessing a well-known
ferromagnetic phase transition \cite{ref21}. For $D/J=3.5$, the
transition width of magnetization does not change until low
frequencies $\omega \leq 5\times 10^{-4}$ with which the width gets
narrower and narrower and $m_{z}(T,\omega)$ curve approaches to the
curve for $D/J=0$. This change apparently corresponds to the change
of the low-temperature peak with frequency in
$\chi^{\prime}(T,\omega)$ shown in Fig. \ref{fig2}. Strikingly,
$\chi_{z}(T,\omega)$ in the inset of Fig. \ref{fig3} exhibits a
sharp peak similar to that of the $D/J=0$ case, i.e., a
ferromagnetic-like MPT. This result supports the coexistence of MPT
and GPT revealed for the case $D/J=4$ of the RMA model in Eq.
(\ref{eq1}) \cite{ref5}. In contrast, the transition in
magnetization of the $D/J=10$ case is quite broad. This is indicated
further in the inset of Fig. \ref{fig3} by the noisy blurring peak
of $\chi_{z}(T,\omega)$ whose height is orders of magnitude lower
than those of $D/J=3.5$ and $D/J=0$ cases. In addition, the
magnetization at low temperatures for the $D/J=3.5$ case is high in
magnitude of 0.7, albeit smaller than 1.0 for the $D/J=0$ case, and
frequency-independent against the small and chaotically
frequency-dependent value of that for the $D/J=10$ case. This
feature is probably due to their different magnetic structures: the
asperomagnet (known in literature as a correlated spin-glass or a
``ferromagnet'' with wandering axis) in the former versus the
speromagnet in the latter \cite{ref1,ref22}. We believe that MPT for
$D/J=10$ likely does not exist or at least is smeared out by strong
RMA and this is why the low-temperature peak is suppressed
completely in $\chi^{\prime}(T,\omega)$ curves.

In general, a phase transition like those for the $D/J=0$ and
$D/J=3.5$ cases has been termed the {\it dynamic transition}, which
is a true thermodynamic phase transition usually studied together
with the {\it dynamic hysteresis} in pure magnetic systems (the
systems without any random defect or anisotropy to pin the magnetic
domains) \cite{ref23,ref24}. These phenomena occur due to a
relaxational delay of the magnetization in response to the, say,
oscillating field. When the oscillation period of the field is much
less than the effective relaxation time of the magnetic system the
hysteresis loop becomes asymmetric about the origin with a
nonvanishing area and a ``{\it spontaneously broken symmetric
phase}" arises dynamically with a nonvanishing value of the dynamic
order parameter $Q$, defined as $Q(T,\omega)=(\omega/2\pi)\oint
m(T,t)dt$ ($Q(T,\omega)$ is the period averaged magnetization and is
equal to $m_{z}(T,\omega)$ in our notation), where the instantaneous
magnetization per site at time $t$ and temperature $T$ is calculated
as $m(T,t)=(1/L^{3})\sum_{i}\vec{S}_{i}(T,t)\cdot\hat{z}$. The
system is in a dynamically-ordered phase when $Q\neq0$ and the loop
is asymmetric or in a dynamically-disordered phase when $Q=0$ and
the loop is symmetric. A transition occurs at $T_{d}$ when one
crosses the boundary separating the two phases. Notice that the
boundary is dynamic in nature since $T_{d}$ depends on both $H_{0}$
and $\omega$, i.e., $T_{d}= T_{d}(H_{0},\omega)$. For any fixed
frequency, the $H_{0}-T$ plane is then divided by the dynamic phase
boundary line $T_{d}(H_{0},\omega)$, which is in general convex
towards the origin. With large values of $H_{0}$, one gets a
``forced oscillation'' kind of scenario inducing the
dynamically-disordered phase ($Q=0$) at high $T$ that gives rise to
low values of $T_{d}(H_{0},\omega)$. All of these features of the
phase diagram have been obtained for pure magnetic systems of Ising
models using mean-field and MC methods \cite{ref24}. They are also
observed in our MC simulation for the weak RMA Heisenberg model as
shown in Fig. \ref{fig4} for the $D/J=3.5$ case with
$H_{0}/J=0.05\sim2.5$ and $\omega=3\times10^{-3}$. Very
interestingly, the dip in $\chi^{\prime}(T,\omega)$ as well as the
only peak in $\chi^{\prime\prime}(T,\omega)$ occur somewhere around
$T_{d}(H_{0},\omega)$ and quite similar to the fashion for two- and
three-dimensional pure Ising models \cite{ref24}. (note, however,
that it is difficult to determine $T_{d}(H_{0},\omega)$ in the
$m_{z}(T,\omega)$ (i.e., $Q(T,\omega)$) curve because
$m_{z}(T,\omega)$ undergoes a gradually broad transition shown in
Fig. \ref{fig4}. Instead, we prefer to take the temperature $T_{a}$
at the peak in $\chi_{z}(T,\omega)$ or equivalently the temperature
$T_{p}$ at the low-temperature peak in $\chi^{\prime}(T,\omega)$ to
construct the diagram of the dynamic transition for the $D/J=3.5$
case shown in the inset of Fig. \ref{fig4}.) Eventually, in ac
susceptibility measurements one may be indicated precisely the same
dynamic transition (where the peaks or dips are shown) as that the
dynamic order parameter (if it could be directly measured in
experiment) provides as long as the values of $H_{0}$ are very small
so that the dynamic transition is continuous because large values of
$H_{0}$ lead to a crossover of continuous/discontinuous transition
at the tricritical point (not shown in our simulation) in the
$H_{0}-T$ diagram \cite{ref23,ref24}. Therefore, the ac
susceptibility for a-Ho$_{28}$Fe$_{72}$ \cite{ref14} is a
particularly prominent example to study experimentally the dynamic
transition in weak RMA systems using the ac susceptibility
measurements.

In summary, our MC simulation shows that the RMA model in Eq.
(\ref{eq1}) can be employed to understand the distinct behaviors in
$\chi^{\prime}(T,\omega)$ of a-Ho$_{28}$Fe$_{72}$ and
Dy$_{40}$Al$_{24}$Co$_{20}$Y$_{11}$Zr$_{5}$ glasses
\cite{ref14,ref15} where the nature of the low-temperature peak of
$\chi^{\prime}(T,\omega)$ for the former is a dynamic transition.
This result marks a striking similarity between weak RMA Heisenberg
model and pure ferromagnetic spin models and sheds light on the
nature of magnetic transition and magnetic ordering, i.e., QLRO, in
magnets with weak random anisotropy.

This work was financially supported by the National Science Council
of Taiwan, R.O.C, under  Grant No. NSC 97-2112-M-007-007-MY3.

\newpage

\newpage

\newpage

\newpage


\begin{references}
 \bibitem{ref1} K. Moorjani and J. M. D. Coey, {\it Magnetic Glasses}, (Elsevier, New York, 1984); D. H. Ryan edt,
 {\it Recent Progress in Random Magnets}, (World Scientific, Singapore, 1992);
 R. W. Cochrane, R. Harris, and M. J. Zuckermann, Phys. Reports {\bf 48}, 1 (1978).
 \bibitem{ref2} Y. Holovatch, V. Blavatoska, M. Dudka, C. Von Ferber, R. Folk, and T. Yavorsokii, Inter. J. Mod. Phys. B {\bf 16}, 4027 (2002); M. Dudka, R. Folk, and Y. Holovatch, J. Magn. Magn. Mater. {\bf 294}, 305 (2005); M. Dudka, R. Folk, Y. Holovatch, and G. Moser,  J. Phys. A {\bf 40}, 8247 (2007).
 \bibitem{ref3}A. Aharony, Phys. Rev. B {\bf 12}, 1038 (1975).
 \bibitem{ref4} U. K. R\"o\ss{}ler, Phys. Rev. B {\bf 59}, 13577 (1999); R. Fisch, Phys. Rev. B {\bf 79}, 214429 (2009).
 \bibitem{ref5} M. Itakura, Phys. Rev. B {\bf 68}, 100405 (2003).
 \bibitem{ref6} Ha M. Nguyen and P. Y. Hsiao, J. Appl. Phys. {\bf 105}, 07E125 (2009).
 \bibitem{ref7} Y. Imry and S. Ma, Phys. Rev. Lett. {\bf 35}, 1399 (1975). \bibitem{ref8} R. A. Pelcovits, E. Pytte, and J. Rudnick, Phys. Rev. Lett. {\bf 40}, 476 (1978).
 \bibitem{ref9} M. Itakura and C. Arakawa, Prog. Theor. Phys. Suppl. {\bf 157}, 136 (2005).
 \bibitem{ref10} N. Avraham., B. Khaykovich, Y. Myasoedov, M. Rappaport, H. Shtrikman, Di. E. Feldman,
 T. Tamegai, P. H. Kesk, M. Lik, M. Konczykowski, K. van der Beek, and E. Zeldov, Nature {\bf 411}, 451 (2001);
 F. F. Bouquet, C. Marcenat, E. Steep, R. Calemczuk, W. K. Kwok, U. Welp, G. W. Crabtree, R. A. Fisher,
  N. E. Phillips, and A. Schilling, Nature {\bf 411}, 448 (2001).
 \bibitem{ref11} D. E. Feldman, Phys. Rev. B {\bf 61}, 382 (2000); Phys. Rev. Lett. {\bf 84}, 4886 (2000).
 \bibitem{ref12} R. Fisch, Phys. Rev. B {\bf 39}, 873 (1989); Phys. Rev. B {\bf 42}, 540 (1990); Phys. Rev. B {\bf 62}, 361 (2000); Phys. Rev. Lett. {\bf 66}, 2041 (1991).
 \bibitem{ref13} R. Harris, M. Plischke, and M. J. Zuckermann, Phys. Rev. Lett. {\bf 31}, 160 (1973).
 \bibitem{ref14}T. Saito, A. Suto, and S. Takenaka, J. Magn. Magn. Mater. {\bf 272-276}, 1319 (2004).
 \bibitem{ref15} Q. Luo, D. Q. Zhao, M. X. Pan, and W. H. Wang, Appl. Phys. Lett. {\bf 92}, 011923 (2008).
 \bibitem{ref16} H. Schremmer and W. Kleemann, Phys. Rev. Lett. {\bf 62}, 1896 (1989).
 \bibitem{ref17}Ha M. Nguyen and P. Y. Hsiao, J. Korean Phys. Soc. {\bf 53}, 2447 (2008).
 \bibitem{ref18}Ha M. Nguyen and P. Y. Hsiao, Appl. Phys. Lett. {\bf 94}, 186101 (2009).
 \bibitem{ref19} K. Binder and A. P. Young, Rev. Mod. Phys. {\bf 58}, 801 (1986).
 \bibitem{ref20}O. V. Billoni, S. A. Cannas, and F. A. Tamarit, Phys. Rev. B {\bf 72}, 104407 (2005).
 \bibitem{ref21} D. P. Landau and K. Binder, {\it A Guide to Monte Carlo Simulations in Statistical Physics},
 (Cambridge Press, Cambridge, Second Edition, 2005).
 \bibitem{ref22}E. M. Chudnovsky and R. A. Serota, Phys. Rev. B {\bf 26}, 2697
 (1982); J. M. D. Coey, J. Appl. Phys. {\bf 49}, 1646 (1978).
 \bibitem{ref23} B. K. Chakrabarti and M. Acharyya, Rev. Mod. Phys.  {\bf 71}, 847 (1999).
\bibitem{ref24} M. Acharyya and B. K. Chakrabarti, Phys. Rev. B  {\bf 52}, 6550 (1995).

  \end{references}
\end{document}